\documentclass[11pt]{article}
\usepackage[T2A]{fontenc}
\usepackage[cp866]{inputenc}
\usepackage[english]{babel}
\usepackage[dvips]{graphicx }
\newcommand{\be}{\begin{equation}}
\newcommand{\ee}{\end{equation}}
\newcommand{\bea}{\begin{eqnarray}}
\newcommand{\eea}{\end{eqnarray}}
\def\vh{\varphi}

\title{Astrophysical Data and Conformal Unified Theory}

\author{ V.N. Pervushin\\
Bogoliubov Laboratory of Theoretical Physics,\\
Joint Institute for Nuclear Research, 141980 Dubna, Russia}

\begin{document}

\maketitle

\begin{abstract}
  Astrophysical data  reformulated in  the units of
  the relative Paris meter and running Planck mass are used for
 restoration of a conformal version of the unified theory where the absolute
 Planck mass belongs to ordinary initial data (like the absolute
 Ptolemaeus  position of the earth belongs to the initial data
 of Newton's mechanics). In the conformal unified theory,
 both the latest data on the Supernova luminosity-distance -- redshift relation
  and primordial nucleosynthesis are described by a free primordial motion of
  the scalar homogeneous field (Scalar Quintessence).
 This primordial cosmic evolution leads to intensive cosmological
 creation of vector bosons forming the baryon asymmetry of the universe
 (during their lifetime) and the CMB radiation as a final product of their decays.
 There are values of the initial data that give the CMB temperature
 (as an integral of the primordial motion) and the energy density budget
 in agreement with observational data.

\end{abstract}

\vspace{3cm}

{\small \bf Invited talk at the International Conference

\vspace{.3cm}

\begin{center}

 ``{\large HADRON STRUCTURE '02}''

\end{center}

\vspace{.3cm}

22 -- 27 September, 2002, in Herlany, Slovakia}

\newpage

\section{Introduction }

The latest astrophysical data on the
 Supernova redshift -- distance relation\cite{snov,sn1997ff}, primordial
 nucleosynthesis,
Cosmic Microwave Background radiation \cite{33}, and baryon asymmetry are
 treated by some specialists  as a revolution in physical cosmology \cite{ch}.

We try to show how these new and old facts of observational cosmology
help us to restore the unified theory and to describe creation of
the universe, time, and matter in agreement with these facts.

\section{Astrophysical Data and Scenarios}

\subsection{Facts}

We restrict ourselves to the following facts of these observations and
measurements.

1. The data on the so-called distance -- redshift relationship testifying to
that the farther cosmic objects, the more  redshift \cite{snov,sn1997ff,f22}
 \be\label{data1}
 z_{\rm cosmic }({\rm coordinate}_c)+1=
 \frac{E_0}{E_{\rm cosmic} ({\rm coordinate}_c)}
 \ee
of spectral lines $E_{\rm cosmic}({\rm coordinate}_c)$ of atoms on these cosmic
objects at the coordinate distance $(r_c={\rm coordinate}_c)$
in  comparison with the present-day spectral $E_0$ lines of the Earth atoms.

2. The data including primordial nucleosynthesis (PN) and the chemical evolution
of the matter in the universe (described in the nice book by Weinberg
\cite{three}) testifying to the definite dynamics of redshift in terms of
measurable time
 \be\label{data2}
 [z_{\rm cosmic} +1]^{-1}|_{({\rm PN})} \simeq \sqrt{({\rm measurable ~~time})}
 \ee
that corresponds to a definite equation of state of matter in the universe.
This equation helps us to determine a kind of matter taking part in the cosmic
evolution of the redshift.

3. The data on the visible number of
particles (baryons, photons, neutrinos, etc.) testifying to that
visible baryon matter gives only the $0.03$ part
 \be \label{b}
 \Omega_{\rm b}=\frac{\rho_{\rm b}}{\rho_{\rm cr}}=0.03
 \ee
 of the critical density $\rho_{\rm cr}$ of
the observational cosmic evolution  \cite{fuk}.

4. The data on the Cosmic Microwave Background radiation with
the temperature $2.7 K$ and its fluctuations \cite{33}.

5. The data on the baryon asymmetry testifying to that the
number of baryons is $10^{-9}$ times less than the number of photons \cite{rub}.

\subsection{Scenarios of the Friedmann-Robertson-Walker Cosmology}

The first scenario (1920-1980) describing observational data was based
on  Einstein's general relativity (GR)
\be\label{gr1}
{\rm GR}[\vh_0|g]=-\int d^4x\sqrt{-g}~\frac{\vh_0^2}{6}~R(g)
\ee
with the Newton constant ${\vh_0^2}/{6}=1/G$
and the measurable interval
\be\label{gr3}
(ds^2)=g_{\mu\nu}dx^{\mu}dx^{\nu}.
\ee
The standard cosmological
model \cite{f22} appeared in the homogeneous approximation of the metrics
\be\label{cffs}
ds^2=(dt)^2-a^2(t)(dx^i)^2
\ee
and began with the {\it anisotropic rigid} Casner state \cite{34}. Then
{\it radiation} appeared in the form of
relativistic  massive particles. When
temperature became lower,
these particles converted into the massive dust {\it matter}.

After 1980, this scenario (anisotropic rigid -- radiation -- dust matter)
was changed by the scenario of
Inflationary Cosmology \cite{linde} (inflation -- radiation -- dust matter) where
 the first state of evolution of the universe is considered
in the form of the law of inflation with a constant $C_I$
\be\label{dataI}
 [z_{\rm cosmic} +1]^{-1}|_{({\rm Beginning})} \simeq
\exp[C_I \times ~t]
 \ee
 to solve some theoretical problems of the old scenario in the way compatible
with the Standard Model (SM) of elementary particles~\cite{db}
identified with excitations of a set of fields ${\rm f}$:
\be\label{sm}
  {\rm SM}~~[M_{\rm Higgs}|{\rm f}]\equiv {\rm SM}~~[y_h~\vh_0|{\rm f}].
 \ee
All masses of elementary
  particles are scaled by the Higgs mass expressed
 in the units of the Planck mass as the absolute parameter of
 general relativity~(\ref{gr1}) in the quantum relativistic region
\be\label{abs}
M_{\rm Planck}=\vh_0 \sqrt{8\pi
\hbar c/3}=2.177\times 10^{-8}{\rm kg},
\ee
 as $M_{\rm Higgs}=y_{\rm h}\vh_0$,~
 $y_{\rm h}\sim 10^{-17}$.

Beginning in 1998, the new data were firstly
obtained \cite{snov,sn1997ff} which made all these
scenarios change. To describe these data in the framework of the
Inflationary Cosmology
with the absolute Paris meter and Planck mass, one needs to
suppose that the $0.7$ part of the energy density of the universe
is in the inflationary state with another very small constant
$C_0<< C_I$ and $0.3$ part is in the state of a dust matter (called the Cold
Dark Matter). The problem appears: how
to explain the origin of all these states
(the primordial inflationary one, radiation, and
the  present-day inflationary state mixing with the Cold Dark Matter)
and their interchanges, and to calculate their energy densities
including the baryon density~(\ref{b})
using the parameters of the
the unified theory (UT) of all interactions in the Riemannian space-time
 \be\label{uf}
 {\rm  UT}[\vh_0|{\rm F}]={\rm GR}[\vh_0|{\rm g}] +
 {\rm SM}[y_h~\vh_0|{\rm F}={\rm g},{\rm f}]
 \ee
and the initial data.

\subsection{Scenarios of the Conformal Cosmology}

To solve this problem, we consider a wider supposition~\cite{plb,039,ppgc}
 that the evolution of the scale of the universe can be determined  not
 only by the theory and initial data, but also by the standard of measurement.

It is worth reminding that the concept of measurable quantities
 in the field theory is no less important than
  the equations of the theory\footnote{''The most important aspect
 of any phenomenon from  mathematical
 point of view  is that of a measurable quantity.
  I shall therefore consider electrical phenomena
  chiefly with a  view to their measurement,
 describing the methods of measurement, and
 defining the  standards on
  which they depend.''
(J.C. Maxwell) \cite{Maxwell}.}.

Suppose that nature selects itself both the theory and
standards of measurement, and the aim of observation is
to reveal not only initial data, but also these measurement standards.
In particular, one of the
central concepts of the modern cosmology is the concept of the scale defined
as the  spatial volume in GR~\cite{bpp}.
If expanding volume of the universe means
the expansion  of ``all its lengths'',
we should specify whether the measurement standard of length
expands.  Here there are two possibilities: the first, the absolute
  measurement standard
\be\label{apm}
 {\rm Absolute~~{}~ Paris~~{}~ Meter} =1 {\rm m}.
 \ee
  does not expand;
  and the second, the relative measurement standard
\be\label{rpm}
 {\rm Relative~{}~~ Paris~{}~~ Meter} =1 {\rm m}\times a(t).
 \ee
 expands together with the universe and
  means that the observable time is identified with the conformal time
 \be\label{cfs3}
d\eta=\frac{dt}{a(t)}.
\ee

Until the present time the first possibility was  mainly considered
in physical cosmology. The second possibility means that
we have no absolute instruments to measure absolute values
 in the universe. We can measure only a ratio of values which does not
 depend on the spatial scale factor. The relative measurement standard
 transforms the spatial scale of the intervals of lengths into
 the scale of all masses
\be\label{rpm1}
 m_c(\eta)=m_f \times a(t)
 \ee
including the Planck one
\be\label{1pct2}
 \vh(\eta)=\vh_0 \times a(t).
 \ee
 which permanently grow.
The spectrum of photons emitted by atoms on far stars  two billion
years ago remembers a size of an atom which is determined by its mass. This
spectrum is compared with spectrum of similar atoms on the Earth
whose mass, at the present time, becomes much larger. This change of the mass
leads to red shift described by the conformal cosmology
defined as the standard cosmology  expressed in terms of the conformal
quantities~\cite{plb,039,ppgc,N}.
 The common point for two cosmologies
is the identification of the evolution of the universe with the evolution
of the same cosmic factor.
In the standard cosmology the cosmic factor scales all distances {\it besides
the Paris meter}~(\ref{apm}). In the conformal cosmology
the cosmic factor scales  all masses~(\ref{rpm1})
{\it including the Planck mass}~(\ref{1pct2}).
As it was shown in a recent paper~\cite{039},
the recent experimental data for
distant supernovae \cite{snov,sn1997ff}, in the case of the
{\it relative} Paris meridian~(\ref{rpm}) correspond to
the cosmic evolution (see Fig.1)
 \be\label{data3}
 [z_{\rm cosmic} +1]^{-1}|_{({\rm Supernova})}(\eta)=\frac{\vh_I}{\vh_0}
  \sqrt{1+2H_I\eta}=\sqrt{1+2H_0(\eta-\eta_0)}.
 \ee
 defined by the initial data: the primordial Planck scale $\vh_I$
 and primordial Hubble parameter $H_I$ related by
 $\vh_I^2 H_I=\vh_0^2 H_0$  with the present-day values of
 the Planck scale $\vh_0$, and the Hubble parameter $H_0$ forming the
 critical density
 $\rho_{\rm cr}=\vh_0^2H^2_0$ at the present-day time $\eta_0$.

\begin{figure}[t]
\vspace{-1cm}
 \begin{center}
\includegraphics[width=0.9\textwidth,clip]{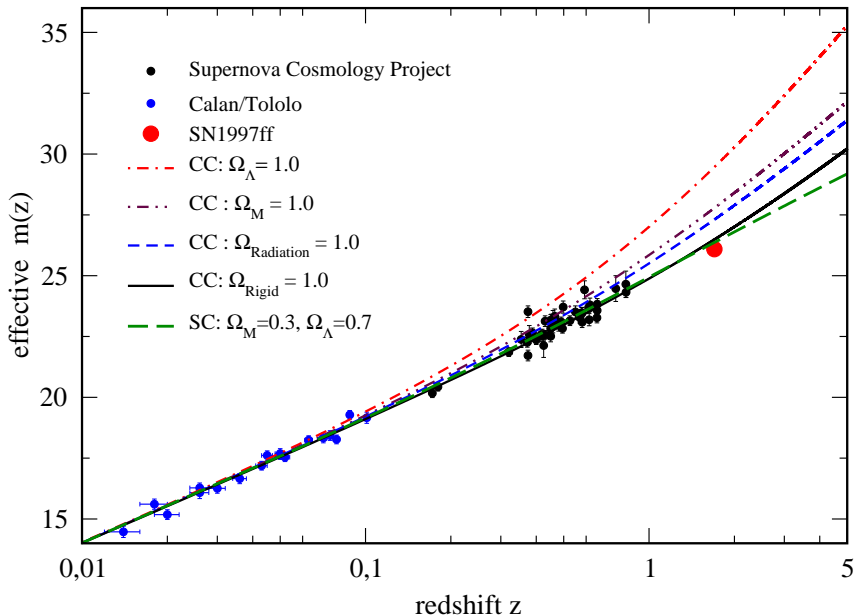}
\caption{{\small The figure taken from~\cite{039} shows the Hubble diagram for
a flat universe model in the
standard cosmology (SC) with the absolute Paris meter~(\ref{apm}) and conformal cosmology
(CC) with the relative Paris meter~(\ref{rpm}).
 The points include  42 high-redshift Type Ia
 supernovae~\protect\cite{snov} and the reported
 farthest supernova SN1997ff~\protect\cite{sn1997ff}. The best
fit to these data  requires a cosmological constant
$\Omega_{\Lambda}=0.7$ in the case of the absolute Paris meter, whereas in
the case of the relative Paris meter
 these data are consistent with  the dominance of the rigid state (\ref{data3}).
 }
\label{fig1}}
\end{center}
\end{figure}

 We see, that, in the case of the relative Paris meter~(\ref{rpm}),
 both the present-day era and the era of the chemical evolution (\ref{data2})
 can be described  by only one isotropic rigid state (\ref{data3}).
 Thus, for the relative Paris meter~(\ref{rpm}) the problem of
 theoretical explanation is simplified, as we have a unique permanent
 state of the cosmic evolution (\ref{data2}) with the
 primordial density
 \be \label{rhoi}
 \rho_{\rm Q}(\eta =0)=\vh_I^2 H_I^2=\frac{\vh_0^2}{\vh_I^2}\rho_{\rm cr}.
 \ee
 The problem is to calculate the densities of the CMB radiation
 $\rho_{\rm CMB}$ with temperature $2.7 K$ (that is a constant
 in the conformal cosmology)
 and visible baryon matter (\ref{b}) from the first principles
 of the unified theory (\ref{uf}) in the relative units~(\ref{rpm}).

\section{Conformal-invariant Unified Theory}

\subsection{Frame symmetry}

The simplification of the cosmic evolution
in the conformal cosmology \cite{plb,039,ppgc} can be considered
as an argument in favor of
that  a group of transformations of
 all possible {\it frames of reference} (i.e., initial data)
is the group of conformal transformations (instead of the Poincare group).

Historically,  frame symmetries appeared as
the Galilean group of transformations rearranging positions and velocities
of initial data of particles in the Newton mechanics, where
the Ptolemaeus absolute position of the earth belongs to
the ordinary initial data.

The frame symmetry of the modern unified theory~(\ref{uf})
is the Poincare group of transformations
rearranging the initial data of relativistic fields.

The frame symmetry of Maxwell's equations is the group of conformal
transformations, as it was shown by Bateman and Cuningham in 1909~\cite{bat}.
 The conformal group was discovered
by M\"obius in the 19th century~\cite{w49}.
The conformal transformations keep invariant the angle between two vectors in
space-time.

 The conformal symmetry of the laws of nature is not compatible with
 absolutes of the kind of
 the  Planck mass as an absolute
 parameter of length, time, and mass in the unified theory  (\ref{uf}).
 The conformal symmetry
 requires converting the unified theory  (\ref{uf})
 into a mathematically equivalent conformal-invariant
 unified theory (CUT)~\cite{plb,039,ppgc},
 where the absolute Planck mass $\vh_0$ belongs to
  the {\it initial data} of a new dynamic
 variable (the dilaton scalar field $w$), like the Ptolemaeus absolute position of the Earth belongs to
 ordinary initial data of the
 dynamic coordinate in Newton's theory.

This, mathematically equivalent to general relativity  (\ref{gr1}),
 conformal-invariant theory of the scalar field $w$ playing the role
 of the absolute measure
 was revealed by Penrose,  Chernikov, and Tagirov (PCT)~\cite{pct}.
 This theory is nothing but the Einstein theory (\ref{gr1})
 for metric multiplied by the square of the scalar field $w$
 \be\label{pct}
  -{\rm PCT}[w|{\rm g}]={\rm GR}~~[1|{\rm g}\times w^2].
 \ee
 In the conformal-invariant unified theory (CUT)
\be\label{cuf}
{\rm  CUT} [w|{\rm g}]=-{\rm PCT}[w|{\rm g}] + {\rm SM}[y_h~w|F]
 \ee
   the PCT action takes the place of the Hilbert one (\ref{gr1}), and the
scalar field $ (y_h~w)$, the Higgs mass in SM $(M_{\rm Higgs} = y_h~\vh_0)$.

The rigid state  (\ref{gr1}) of the Supernova cosmic evolution can be
explained by the free homogeneous motion of the Scalar Quintessence  (SQ) \cite{ppgc}.
 Both the running Planck
 mass $w$ and Quintessence $Q$ are described by the difference of two
 Penrose-Chernikov-Tagirov actions
 (\ref{pct}) ${\rm PCT}[w_-|{\rm g}]-{\rm PCT}[w_+|{\rm g}]$, where
 $w_+=w\cosh Q$, $w_-=w\sinh Q$ \cite{ppgc}. Finally, the conformal-invariant
 unified theory of all interaction takes the form
\be\label{cuf2}
{\rm  CUT} [w|{\rm g}]=-{\rm PCT}[w|{\rm g}] + {\rm SQ}[w|{\rm g}]
+{\rm SM}[y_h~w|F],
 \ee
where
\be\label{SQ}
 {\rm SQ}[w|{\rm g}]=\int d^4x \sqrt{-g}~w^2~\partial_\mu Q\partial^\mu Q.
 \ee

\subsection{Frame-fixing formulation}

The old unified theory~(\ref{uf}) (supplemented by the Quintessence)
 with an absolute Planck scale of mass
 appears as the
 Ptolemaeus choice of a frame of reference
 \be\label{pct1}
 w_f=\vh_0={\rm constant},
 \ee
 with the absolute measurement standard  (\ref{apm}).
One of the main cosmological consequences of the absolute choice
 is expanding
 the  spatial volume of the universe
 $V_f=V_f(t)$ and all lengths in
 the universe with respect to the absolute Paris meter (\ref{apm}).
 Paris is a nice place in the universe. But why is
 the meter defined in 1791 as a $1/40,000,000$ part of Paris
 meridian so distinguished?
In any case, we can include into  {\it all lengths}
 the measurement standard itself and use  the relative Paris meter (\ref{rpm}).
  Then our measurements of {\it all lengths} as ratios are not expanding.
  The measurable spatial volume of the universe is a constant $V_c$.
 While, the measurable Planck mass in the same theory becomes a dynamic variable
 (\ref{1pct2}) with all measurable masses of elementary particles (\ref{rpm1}).

 The relative Paris meter (\ref{rpm}) means the Copernicus choice of
 a frame of reference in the field space
 \be\label{pct2}
 w_c=\vh(\eta), ~~~~V_c={\rm constant}
 \ee
 instead of the Ptolemaeus (\ref{pct1}) with the running volume $V(t)$.

 Thus, {\it the relative Paris meter} (\ref{rpm}) {\it converted
 the Planck absolute mass into ordinary initial data,
 like Copernicus and Newton  converted
 the Ptolemaeus absolute position of the Earth into ordinary initial data.}

In 1974, Barbashov and Chernikov \cite{bc}
applied the same frame-fixing formulation to the  relativistic string
theory
and proved that this theory coincided with the Born-Infeld theory
that strongly differs from the abstract frame-free formulation
of a string with the so-called  Virasoro algebra~\cite{bn}.
Reiman and Faddeev~\cite{rf}
reproduced and generalized this result in 1975 (for details see~\cite{bpp}).
 Both
the first Dirac quantization of electrodynamics~\cite{Dir} and
 the Schwinger quantization of modern gauge theory \cite{sch2,mpl}
were fulfilled in a concrete frame.
As it was shown by Schwinger
 \cite{sch2,mpl}, this description does not contradict the general theory
 of irreducible and unitary representations of the relativistic group
 constructed  by I.M. Gel'fand and M.I. Graev, and
 V. Bargmann, E.P. Wigner, and A.S. Wightman (see the monographs
\cite{Schweber}, \cite{Logunov}).

In the frame-fixing formulation,
the problem of energy and time of the universe
 is solved by the identification of
 the measurable energy of the universe $E$ with respect to
 the evolution parameter~$\vh$
 with the canonical momentum of this parameter~\cite{plb,bpp,pp}
 $P_\vh=\pm E$.
 This energy of the universe is  similar to the
 frame-dependent definition of  energy  for
 a relativistic particle in special relativity
   by Poincare and  Einstein~\cite{poi}, and
 for a relativistic string, by Barbashov and Chernikov~\cite{bc,rf,bpp},
 who got the Born-Infeld model instead of the Virasoro algebra.

\subsection{Geometrization of the energy constraint}

To explain  the world by solving the Newton equation Laplace required initial data
for the position and velocity of all particles in the absolute space-time.
To obtain  these data Laplace had to fix a frame of reference.
If  modern Laplace tries to explain a relativistic
world fixing  a frame like Poincare and Einstein \cite{poi},
he will lose the dynamics of the particle in the world Minkowski space $[X_0|X_i]$
with respect to the geometric (i.e., proper) time-interval $\eta$, together
with the pure relativistic effects of
 the type of relativistic time dilatation.
This dilatation means that
 an unstable particle in the rest frame of an observer
 has greater lifetime than its
geometric lifetime in its comoving frame.
If Laplace tries to reduce a relativistic system
to the Newton one, he loses pure relativistic effects.

The loss of the geometric time interval after fixation of a frame
is the common problem of all relativistic theories including
relativistic cosmology of the universe.
Fixing a concrete frame Poincare and Einstein
lost the proper {\it time-interval} in Special Relativity
(where the role of the time-like evolution parameter is played by
the definite variable $X_0$ in the world space $[X_0|X_i]$) and Wheeler
and DeWitt \cite{M} lost the {\it time-interval} in General Relativity
(where the role of the time-like evolution parameter is played by
the scale factor, i.e., the running Planck mass  $\vh$ in the world space
 $[\vh|F]$).

In the gauge constrained relativistic
theory one could obtain only the dependence of
fields $[F]$ on this field evolution parameter $\vh$:
$[F(\vh)]$ with the initial data at $\vh=\vh_I$.
These initial data $[\vh_I|F(\vh_I)]$ at the world field space
are treated as the field coordinates of a point of the greatest event --
 the  creation of the universe in this world field space out of the time.
Having these initial data one can describe
 the wavefunction of the universe $\Psi_F[\vh \geq \vh_I|F,F_I)]$ as the amplitude
of the probability to find the universe at the point $[\vh|F]$, if
it was  created at the point $[\vh_I|F_I]$,
 in the world field space out of the {\it time}.
The {\it geometric time} is a superfluous element in both special relativity
with the initial data in the Minkowski space-time and
 general relativity with the initial data in the field space.

 In the case of a relativistic particle, there are two
methods to  obtain its dynamics with respect to the geometric time-interval:
to change the frame, or to introduce one more reality (called geometric)
of a particle in the same frame with the time-interval constructed by
  {\it straightening} the gauge constraint \cite{bpp,pp}.
The second method belongs to the Italian mathematician Levi-Civita
who supposed to straighten a constraint in the theory of
differential equations as far back as 1906 \cite{lc}. The Levi-Civita  method
of geometrization of the Minkowski world space means a transition
to the geometric variables
\be\label{lcp}
[X_0|X_i]~~ \Longrightarrow~~[Q_0|Q_i],
\ee
where one of the variables coincides with the time-interval: $Q_0=\eta$.
A relativistic
particle can be completely described by two Newton-like {\it realities} in two world
spaces: X-space and Q-space (\ref{lcp})
supplemented by their {\it relationship} in the form
of the  {\it geometrization} (\ref{lcp})
\be\label{lcp1}
X_0(\eta,Q_i),~~~~X_i(\eta,Q_i).
\ee
In this case, one
should require two sets of the initial data in each reality. Then we
obtain two wavefunctions $\Psi_X[X_{0}\geq X_{0I}|X_i,X_{iI}]$ and
$\Psi_Q[\eta \geq 0|Q_i,Q_{iI}]$. The relationship (\ref{lcp1})
is treated as a new, in principle, element of the scientific explanation
of the pure relativistic effects.

This {\it geometrization} \cite{lc} of the energy constraint is the universal
method of a consistent description of all relativistic  systems including
a string \cite{bpp} and a universe \cite{pp}.
In the conformal-invariant unified theory (\ref{cuf}),  (\ref{pct2})
 this method converts the field space
\be \label{fv}
 [\vh~|~F={\rm Field~ variables}]
 \ee
with the field evolution parameter $\vh$ into the geometric world space
 \be \label{gv}
 [\eta~|~G={\rm Geometric~variables}]
 \ee
 with the time evolution parameter \cite{pp}.
The  {\it geometrization}  as a rigorous mathematical construction
of  the geometric time $\eta$ includes the
transformations of the initial fields $F$ into the
  geometric fields $G$ (known as the Bogoliubov transformations)
and introduces
into the theory the cosmic initial data $G_I$ at the beginning of
the universe $\eta=0$
(including a number of particles of geometric fields) \cite{bpp,pp,ps1}.

The universe like a relativistic particle has
  two  {\it realities}: field
 and geometric. Each of them has its world space of variables
((\ref{fv})
or (\ref{gv})), its  evolution parameter
(the cosmic scale factor $\vh$ or geometric time $\eta$),
and its wavefunction (the field $\Psi_F[\vh\geq \vh_I|F,F_I]$ or  geometric
$\Psi_G[\eta \geq 0|G,G_0]$).
 The evolution of cosmic scale factor  with respect to time
 \be \label{vt}
 \vh(\eta)
 \ee
  is considered as a pure relativistic effect of the
geometrization of the energy constraint that is beyond the scope of
 the Newton-like mechanics.

Thus, the {\it absolute-free} conformal symmetry
of the unified theory, its {\it concrete} frame-fixing
fundamental formulation, and the geometrization of the energy constraint
with two {\it realities} of the universe in the field space and the geometric one
make up a new framework of explanation of all physical facts
including physical cosmology. And this explanation should be considered
on equal footing with the old scheme keeping the Newton absolutes of the
type of the absolute Paris meter (\ref{apm}) and the absolute Planck era.

\section{``Big Bang'' as creation of the universe}

\subsection{Creation of the universe and time}

The wavefunction of the universe in the field reality
\be\label{field}
\Psi_{\rm field}[\vh,\vh_I|Q,Q_I;F,F_I]=
\ee
$$
A^+_E \Psi_{\rm universe}[\vh \geq \vh_I|Q,Q_I;F,F_I]+
A^-_E \Psi_{\rm anti-universe}[\vh \leq \vh_I|Q,Q_I;F,F_I]
$$
describes the greatest events --
the creation of the universe with  positive energy
without the cosmic singularity $\vh \geq \vh_I$, or the annihilation
of the anti-universe also with positive energy and
the cosmic singularity $\vh \leq \vh_I$.
To make this creation  stable, one should to
 construct  the wavefunction of the quantum universe in the field reality
excluding the negative value of the energy $P_\vh=-E$ from the wave function.
To do so one needed to treat the creation of the universe with negative energy as
annihilation of the anti-universe with positive energy.
This construction
is known in quantum field theory as causal quantization with the operators of
 creation $A^+$ and annihilation $A^-$ of the
 universe~\cite{Schweber,Logunov}.
Consequences of the causal quantization are the positive arrow
 of the geometric time and its beginning~\cite{bpp,pp} $\eta\geq 0$.

The wavefunction of the universe in the geometric reality
(including the Quintessence $Q_g$)
\be\label{geom}
\Psi_{\rm geometric}[\eta\geq 0|Q_g,G]
\ee
describes the quantum evolution of the universe in the geometric
world space $[\eta|Q_g,G]$ with  initial data for matter fields.
The primordial Hubble parameter $H_I$ sets  a natural unit of time $\eta_I=1/2H_I$.
We can choose the zero initial data for matter
fields,  as a stable state with the lowest energy, i.e., the quantum field vacuum.

\subsection{Creation of matter}

In the inflationary models~\cite{linde} it is proposed that
 from the very beginning the universe is a hot fireball of massless particles
 that undergo a set of phase transitions. However, the origin of particles
 is an open question as the isotropic evolution of the
 universe cannot create massless particles \cite{tag}.
 Nowadays, it is evident that
 the problem of the cosmological creation of
 matter from vacuum is beyond the scope of the inflationary model.

 Here we list arguments in favor of that
 the cosmological particle creation from vacuum
 in the conformal-invariant unified theory can describe
  the cosmic energy density budget of observational cosmology.

At the first moment $\eta_I=1/2H_I$
of the lifetime of the universe, the fundamental frame-fixing quantization
\cite{sch2,mpl} of
 the modern version of the unified theory
shows us an effect of the intensive cosmological creation \cite{tag} from
the geometric Bogoliubov vacuum
of relativistic massive vector
bosons~\cite{039,ppgc,114}\footnote{The effect of intensive creation of bosons
 is a pure relativistic consequence of the fundamental operator
 quantization \cite{sch2,mpl}, as
a change of the field variables removing this effect violates relativistic
covariance,
i.e., violates the Poincare algebra of observables constructed from these
variables.}.
The distribution functions of the longitudinal
$ {\cal N} ^ {||}$ and
transverse $ {\cal N} ^ {\bot}$ vector bosons  calculated in~\cite{039,ppgc,114}
for the initial data $H_I=M_I $
are introduced in Fig. 2.

The choice of the initial data $ M_v(\eta=0)=M_{I}$ is
determined by the lower
 boundary for a boson mass from the area of its initial values
 allowed by the uncertainty  principle $ \delta E ~\eta_I\geq \hbar $
for energy variations of energy $ \delta E=2M_I $
at creation of a pair of bosons in the universe with
 minimum lifetime for the  case considered $ \eta_I=1/2H_I $.
We can speak about the cosmological creation of a pair
of  massive particles in the universe,  when the particle mass
$ M_v(\eta=0)=M_{I}$  is larger than the primordial Hubble parameter
$M_{I} \geq H_I $.

The distribution functions of the longitudinal
$ {\cal N} ^ {||} (x, \tau) $  vector bosons
introduced in Fig. 2 show the large contribution of relativistic momenta.
This means the relativistic dependence of the particle density on the temperature
in the form $n(T)\sim T^3$. These distribution functions show also
that the time of establishment of the density and temperature
is the order of the inverse primordial Hubble parameter. In this case,
one can estimate  the temperature $T$ from the equation
in the kinetic theory \cite{ber} for the time of establishment of the temperature
$$
\eta^{-1}_{relaxation}\sim n(T)\times \sigma \sim H,
$$
where $\sigma \sim 1/M^2$ is the cross-section.

This kinetic equation and values of the initial data $M_I = H_I$
give the temperature of relativistic bosons~\cite{plb,039,ppgc,114}
$$
 T\sim (M_I^2H_I)^{1/3}=(M_0^2H_0)^{1/3}=2.7 K
$$
as a conserved number of cosmic evolution compatible with
the Supernova data \cite{snov,sn1997ff} and the primordial chemical evolution
\cite{three}.
 We see that
this calculation gives the value surprisingly close
to the observed temperature of the CMB radiation $ T=T_{\rm CMB}= 2.73~{\rm K}$.

\begin{figure}[t]
 \includegraphics[width=0.57\textwidth,height=0.51\textwidth,angle=-90]{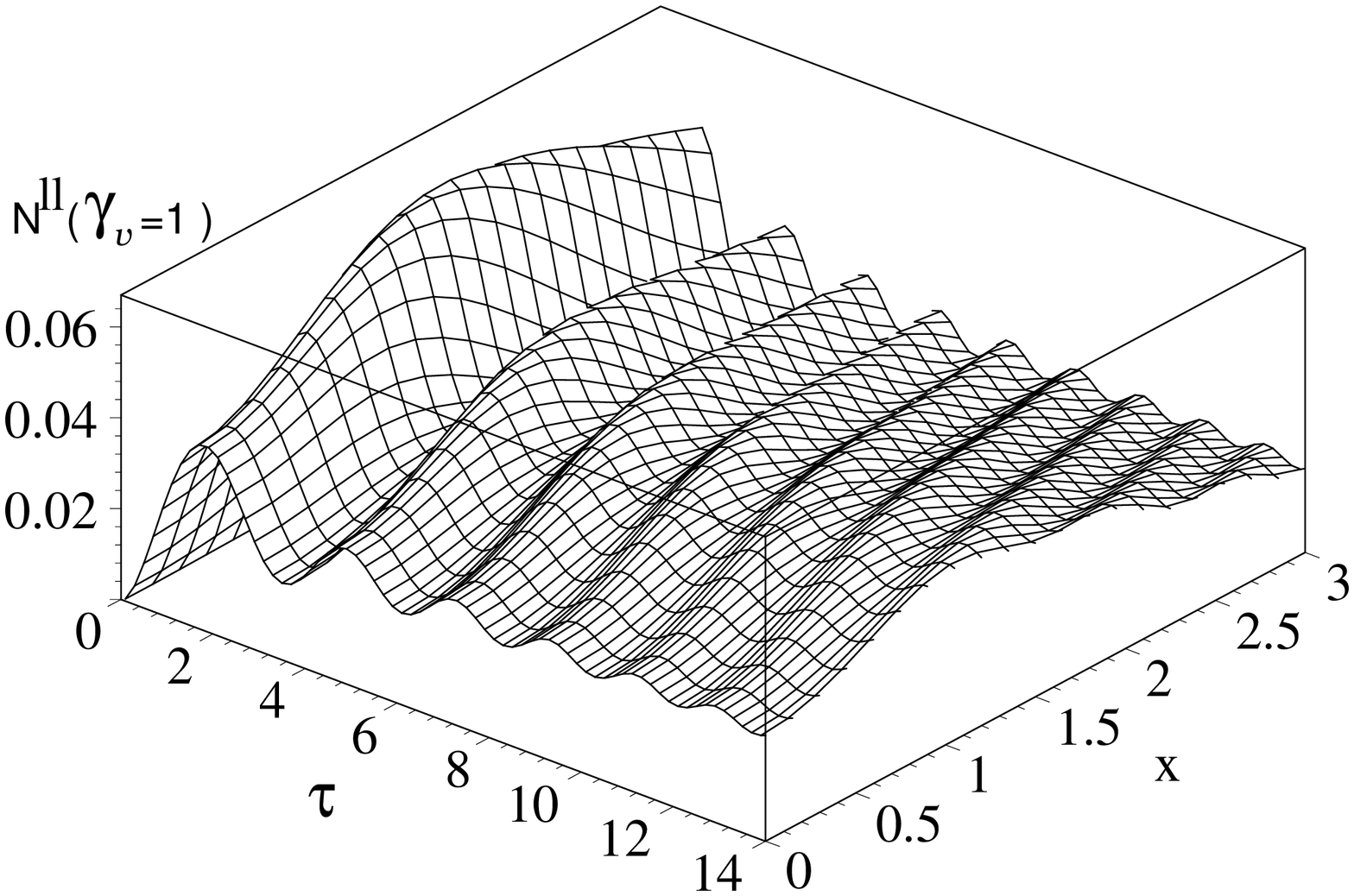}\hspace{-5mm}
 \includegraphics[width=0.57\textwidth,height=0.51\textwidth,angle=-90]{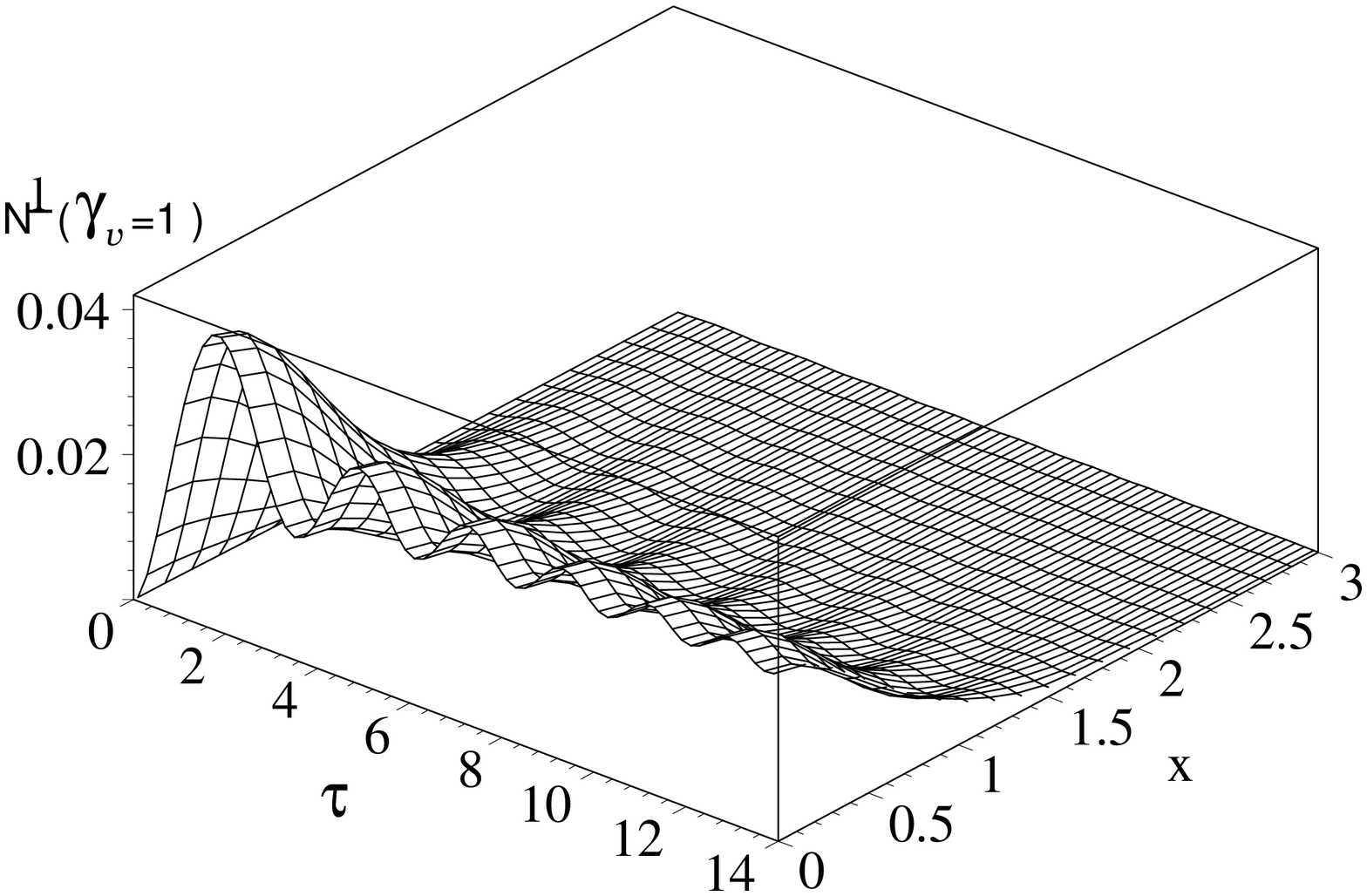}
  \caption{\small The figure taken from \cite{114} shows dependence of
longitudinal $N^\|$ and transverse $N^\bot $  components of the
distribution function of vector
bosons describing their very fast creation in  units of the primordial lifetime
of the universe $ \tau=2H_I \eta $ \cite{ppgc,114}.
Their momentum distributions in  units of the primordial mass $ x = q/M_I $ show
the large contribution of longitudinal bosons and their
relativistic nature.
}
\end{figure}

A ratio of the density
of the created matter
$ \rho_{\rm v}(\eta_I)\sim T^4$ to the
density of the primordial cosmological motion of the universe
$ \rho_{\rm cr.}(\eta)=H_I^2\vh^2_{I}$ has an
extremely small number
\be\label{data5}
\frac{\rho_{\rm v}(\eta_I)}{\rho_{\rm Q}(\eta_I)}\sim
\frac{M^2_{I}}{\vh_I^2}=\frac{M^2_{W}}{\vh_0^2}
\sim 10^{-34}.
\ee

 On the other hand, it is possible to estimate the
 lifetime of the created bosons in the early universe
 in dimensionless unites
$\tau_L= \eta_L/\eta_I $, where $ \eta_I = (2H_I) ^ {-1} $, by
utillizing an equation of state $\vh^2 (\eta_L) =\vh_I^2 (1 +\tau_L) $ and
define the lifetime of $W$-bosons in the Standard Model
\be \label{lifes}
1+\tau_L=
\frac{2H_I\sin^2 \theta_{\rm W}}{\alpha_{\rm QED} M_W(\eta_L)}=
\frac{2\sin^2 \theta_{\rm W}}{\alpha_{\rm QED}\sqrt{1+\tau_L}}~,
\ee
where $\theta_{\rm W}$ is the Weinberg angle,  $\alpha_{\rm QED}=1/137$.
The solution of equation~(\ref{life}) gives the value for $M_{vI}\simeq H_I$
\be \label{life}
\tau_L+1=
\left(\frac{2\sin^2\theta_{\rm W}}{\alpha_{\rm QED}}\right)^{2/3}
\simeq {16}.
\ee
The transverse bosons during their lifetime
form the baryon symmetry of the universe
 as a consequence of the ``polarization'' of the
Dirac sea vacuum of left fermions by these bosons,
 according to the selection rules of the
Standard Model~\cite{ufn} with
left current interaction in SM
$j_{L\mu}^{(i)}=\bar \psi^{(i)}_L\gamma_\mu\psi^{(i)}_L$
for each left doublet $\psi^{(i)}_L$ marked by
an index $(i)$.
At a quantum level, we have an abnormal current
\bea \label{ac}
\partial_\mu j_{L\mu}^{(i)}=-\frac{{\rm
Tr}\hat{F}_{\mu\nu}{}^*\!{\hat{F}_{\mu\nu}}}{16\pi^2},
\eea
where in the lowest order of perturbation theory (p.t.)
$\hat{F}_{\mu\nu}=
-({\imath g\tau_a}/{2})(\partial_\mu v_\nu^a-\partial_\nu v_\mu^a)$.

Integration of the equation (\ref{ac}) gives the number of left fermions
$N=\int d^4x \sqrt{-g}\partial_\mu j_{L\mu}^{(i)}$ created
during the lifetime $\eta_L=\tau_L\times\eta_I$ of vector bosons
\bea \nonumber
 N(\eta_L)=
 -\int_0^{\eta_L} d\eta \int \frac{d^3 x}{16\pi^2} \;
 {}_{\rm sq}\langle 0|{\rm Tr}\hat{F}_{\mu\nu}
 {}^*\!{\hat{F}_{\mu\nu}}|0\rangle{}_{\rm sq} \equiv N_{\rm W}+N_{\rm Z},
\eea
where
$$ 
 N_W =
\frac{4{\alpha}_{\rm QED}}{\sin^{2}\theta_{\rm W}}\int_{0}^{\eta^{W}_{\rm l}}
d\eta \int \frac{d^3 x}{4\pi}~~{}_{\rm sq}
\langle 0|E^{W}_{j}B^{W}_{j}|0\rangle{}_{\rm sq}~,
$$
$$
 N_Z =
\frac{{\alpha}_{\rm QED}}{\sin^{2}\theta_{\rm W}\cos^{2}\theta_{\rm W}}
\int_{0}^{\eta^{\rm Z}_{\rm l}} d\eta \int
\frac{d^3 x}{4\pi}~~{}_{\rm sq}
\langle 0|E^{Z}_{j}B^{Z}_{j}|0\rangle{}_{\rm sq}~.
$$
$$
\int  \frac{d^3 x}{4\pi}~{}_{\rm sq}
\langle 0|E^{v}_{j}B^{v}_{j}|0\rangle{}_{\rm sq}
= -\frac{V_0}{2} \int\limits_{0 }^{\infty }dk |k|^3
\cos(2\theta_v)\sinh(2r_v)~,
$$
$\theta_v$ and $r_v$ are given by the Bogoliubov equations \cite{ppgc,114}.
The lifetime of bosons
$ \tau_L^W = 15 $, $ \tau_L^Z = 30 $,
 leads to estimation of the magnitude of the nonconservation of fermion number
\begin{eqnarray}\nonumber
\Delta F&=&\frac{( N_W+ N_Z)}{V_0}
\\ \nonumber
 &=&\frac{{\alpha}_{\rm QED}}{\sin^{2}\theta_{\rm W}}
  { T^3 }
  \left(4\times 1.44+\frac{2.41}{\cos^{2}\theta_{\rm W}}\right)
  =1.2  n_{\gamma}~,
\end{eqnarray}
where $n_{\gamma}=({2.402}/{\pi^2 })T^3$ is the density of number of the CMB photons.
The baryon asymmetry appears as a consequence of three Sakharov
conditions:
${\rm CP}$-nonconservation, evolution of the  universe $H_0\not= 0$
and the violation of the baryon number~\cite{114}
$$
{\Delta B}=X_{\rm CP}\frac{\Delta F}{3}=0.4X_{\rm CP}n_{\gamma}~,
$$
where $X_{CP}$ is a factor determined by a superweak interaction
of $d$ and $s$-quarks $(d+s~\rightarrow ~s+d)$
with the CP-violation experimentally observed in decays of
$K$ mesons with  a constant of a weak coupling
   $X_{\rm CP}\sim  10^{-9}$~\cite{o}.

After the decay of bosons, their temperature is inherited by the
Cosmic Microwave Background  radiation.
All the subsequent evolution of matter with varying masses
in the constant Universe replicates  the
well-known scenario of the hot
universe~\cite{three}, as this evolution is determined
by the conformal-invariant ratios of masses and temperature $m/T$.

As the baryon density increases as a mass and the Quintessence density decreases
as the inverse square mass, the present-day value of the baryon density can be
estimated by the relation
\be\label{data7}
\Omega_{\rm b}(\eta_0)=\left[\frac{\vh_0}{\vh_L}\right]^3
\frac{\rho_{\rm b}(\eta_{\rm L})}{\rho_{\rm Q}(\eta_{\rm L})}=
\left[\frac{\eta_I}{\eta_L}\right]^{3/2}
\sim \left[\frac{\alpha_{\rm QED}}{\sin^2 \theta_{\rm W}}\right]
\sim 0.03,
\ee
if the baryon
asymmetry with the density
\be\label{data6}
\rho_{\rm b}(\eta=\eta_{\rm L})
\simeq 10^{-9} 10^{-34}\rho_{\rm Q}(\eta=\eta_{\rm L})
\ee
was frozen by the superweak interaction. This estimation gives the value
surprisingly close to the observational density (\ref{b})
in the agreement with the observational data.
There are arguments \cite{mar} in favor of
that  cosmological creation of particles shown on Fig. 2
can also describe the primordial fluctuations of temperature of CMB \cite{33}.
Generally speaking,  all these present and future results can only be treated as a set of
arguments in favor of the considered unified theory.

\section{Conclusion}

Thus, we have shown that the conformal-invariant unified theory (\ref{cuf2}) with
geometrization of constraint and frame-fixing with
the primordial initial data $\vh_I=10^{4} {\rm GeV}$,
$H_I= 2.7~ {\rm K}=10^{29}H_0$
(determined by a free homogeneous motion of
the Scalar Quintessence, i.e., its electric tension)
can describe the following
events:
$$
\begin{array}{rll}
&\eta =0 &{\rm creation~ of~ the ~ ``empty"~universe~from ~``nothing"}\\ [1.5mm]
&\eta \sim 10^{-12}s &{\rm creation~ of~ vector~bosons~from ~``nothing"}\\ [1.5mm]
10^{-12}s <\!\!\!\!&\eta < 10^{-11}\div 10^{-10} s~~~&{\rm formation~ of~ baryon~asymmetry}\\ [1.5mm]
&\eta \sim 10^{-10}s&{\rm decays~ of~ vector~bosons}\\ [1.5mm]
10^{-10}s <\!\!\!\!&\eta < 10^{11}s&{\rm primordial~ chemical~ evolution ~of~matter}\\ [1.5mm]
&\eta \sim  10^{11}s&{\rm recombination,~or~separation ~of ~CMB}\\ [1.5mm]
&\eta \sim  10^{15}s&{\rm formation~of~galaxies }\\ [1.5mm]
10^{17}s <\!\!\!\!&\eta&{\rm hep~experiments~and~Supernova ~evolution}.
\end{array}
$$

In this case, coordinates of
the point of creation of the universe in the world field space $\vh_I=\vh_0/(z_I+1)$
are considered as ordinary initial data.
The absolute Planck mass (\ref{abs}) is determined by the
present-day values of these field coordinates, like
the absolute Ptolemaeus  position of the earth is determined by the
 present-day values of its spatial coordinates in the Newton theory.

\vspace{0.2cm}

The author is  grateful to Profs. B.M. Barbashov, N.A. Chernikov,
P. Flin, and J. Lukierski
for fruitful discussions.
This work was supported by
the Bogoliubov-Infeld Programme.

\end{document}